\documentstyle[aps,epsf]{revtex}  % DO NOT DELETE THIS LINE 

\begin{document}        %  DO NOT DELETE OR CHANGE THIS LINE

\def \cl {{\cal L}}
\def \be {\begin{equation}}
\def \ee {\end{equation}}
\def \bea {\begin{eqnarray}}
\def \eea {\end{eqnarray}}
\def \bi {\bibitem}
\def \ci {\cite}
\def \e {{\rm e}}
\def \o {\omega}
\def \a {\alpha}
\def \n {\nu}
\def \cf {{\cal F}}
\def \L {\Lambda}
\def \x {\xi}
\def \g {\gamma}
\def \del {\partial}
\def \pt {{\rm PT}}
\def \eps {\varepsilon}
\newcommand\re[1]{(\ref{#1})}
\def \lab #1 {\label{#1}}

\def \p {\pi}
\def \m {\mu}
\def \cs {{\cal S}}
\def \as {{\alpha_s}}

\baselineskip 14pt
\title{\ }
\begin{flushright}
ITP-SB-99-22
\end{flushright}

\bigskip

\centerline{\Large {\bf Recent Progress in QCD}\footnote{Plenary talk presented at 
the meeting of the American Physical Society 
Division of Particles and Fields (DPF 99), UCLA, Los Angeles, CA, 5-9 Jan 1999.}}

\medskip
\centerline{George Sterman} 

\smallskip

\centerline{\it Institute for Theoretical Physics, SUNY at Stony Brook, Stony Brook,
NY 11794-3840} 
%\maketitle              % Creates the title area, Do Not Remove

\begin{abstract}        % Do Not Delete this line
The past few years have seen remarkable progress in the theory and phenomenology
of QCD, bringing perturbative and nonperturbative methods
into closer contact with each other and with experiment.  
\end{abstract}   	% Do Not Delete this line

\section{Introduction: QCD a Group Portrait}               % Introduction goes below.

In this talk, I will summarize some recent developments in QCD,
concentrating on, but not limited to, topics discussed at
this meeting.  Details,
of course, can be found in the talks themselves, presented in 
lively sessions organized by Lance Dixon and Joey Huston
for perturbative QCD, and by Paul Mackenzie and Claude Bernard
for nonperturbative QCD. The future of the field is
in the already-advanced convergence of these topics,
sometimes thought of as nearly independent.  
Progress on heavy quark physics is described in the
proceedings from separate parallel sessions and plenary talks.
I must necessarily pass over some of the most interesting
recent advances in other closely related fields as well for lack of space.

I have tried in the following to interleave perturbative and
nonperturbative treatments of QCD dynamics.  Let me begin
with a few general comments on the place of QCD studies
in high energy physics.

{\it Why QCD?}  By now it is clear that QCD is a ``correct", or phenomenologically relevant
theory, at least the way classical Maxwell theory is
``correct".  Classical electromagnetism is an effective
theory appropriate to the limit of many photons; quantum chromodynamics might
be the long-distance limit of some more fundamental, underlying theory.
Its self-consistency, however, leaves us to free to study QCD in its own terms.

This is a fascinating, and dauting, task, despite the fact that 
QCD is defined through a
 single dimensionless parameter, $\alpha_s$.    The overall dimensional
scale is determined by comparison with other interactions, 
for example, by measuring the
strong coupling at the mass of the Z: $\alpha_s(M_{\rm Z})$.
Its intrinsic interest aside, QCD has important ``practical" applications,
in the calculation of backgrounds to new physics
and, for hadron colliders particularly, in predictions for new 
particle cross sections.
From a theoretical point of view, however, what makes QCD so
attractive is that it is a quantum
field theory that requires all orders in perturbation theory and
nonperturbative analysis to confront data at available
energies.  

{\it QCD is the exemplary quantum field theory.}  QCD exhibits
most of the classic quantum field-theoretic phenomena discovered
in the sixties and seventies, including asymptotic freedom, 
confinement, spontaneous symmetry breaking and instantons.
The problems of strong interactions that
gave rise to QCD were also, in the same time
period, the original inspiration for concepts of duality and strings.
As we saw at this conference,
the strong interactions, now understood as QCD,
are once more the meeting ground for field theory, duality and string
theory.

Again, QCD is evidently the correct theory of the strong interactions.
Given its depth, however, ``tests" 
of QCD should be thought of as tests, or perhaps better,
``explorations", of quantum field theory itself.   Complementary to the
extraordinary accuracy of selected perturbative predictions in quantum electrodynamics
are the broad predictions of QCD, interweaving nonperturbative and
perturbative scales and phenomena.

\section{QCD at the Shortest Distances}

It was the asymptotic freedom of QCD that first drew attention
to gauge field theory as the unique description of the strong
interactions at short distances.  This theme continues to unfold
in current experiments, and to provide a basis for extrapolations
between energy scales and the detection of signals for new physics.
Let us begin with a run-through of the underlying methods
\footnote{I have described some of the technical background
in Ref\ {\protect \cite{tasi95}}.}.

\subsection{Methods}

{\it Infrared safety.} Infrared safe quantities are
insensitive to long-distance effects, and may be calculated
in perturbation theory \cite{tasi95,glover}.  An infrared safe 
quantity may or may not be directly observable.
The classic examples include
the total cross section for $\e^+\e^-$ annihilation to hadrons,
and jet and event shape cross sections in $\e^+\e^-$.  These can
be written in the general form, 
\be
Q^2\; \hat \sigma_{\rm phys}(Q^2)
=
\sum_n c_n(Q^2/\mu^2)\; \as^n(\mu)\, ,
\label{IRS}
\ee
with the $c_n$ dimensionless functions of the 
ratio of the hard scale to the renormalization scale $\mu$.

{\it Factorization.} Cross sections for
deep-inelastic scattering (DIS) and for jet or heavy quark production in hadron-hadron
scattering, are
not purely perturbative, but appear as convolutions
in parton momentum fractions of distributions
$f_{a/h}(\xi,\m)$ of partons $a$ in
hadrons $h$, with perturbative hard-scattering functions $\hat\sigma^{\rm PT}$,
\be
Q^2\sigma_{\rm phys}(Q,x)
=
\sum_{{\rm partons}\ a}\hat \sigma_a^{\rm PT}(Q/\mu,\as(\mu))\otimes f_{a/h}(\mu)
= 
\sum_{{\rm partons}\ a}
\int_x^1 d\xi\;  \hat\sigma_a^{\rm PT}(x/\xi,Q/\mu,\as(\mu)) f_{a/h}(\xi,\mu)\, ,
\label{fact}
\ee
where in the second equality we have exhibited the convolution
appropriate to deeply inelastic scattering (with $x=Q^2/2p\cdot q$).
Corrections to Eq.\ (\ref{fact}) are suppressed by ${\cal O}(1/Q^2)$ \cite{tasi95}.
In  this formula, there is a complementarity between
the roles of parton distributions and the hard scattering.
The distributions $f_{a/h}$ are universal among hard-scattering
processes, but particular to hadron $h$, while the 
functions $\hat \sigma^{\rm PT}$ are particular to the process,
but universal  among external hadrons.  This last feature
enables us to calculate realistic $\hat\sigma^{\rm PT}$ in ``unrealistic", but
technically manageable (infrared regulated) scattering processes, in which
the initial state hadrons are partons.

{\it Evolution.} The physical cross sections of Eqs.\ (\ref{IRS})
and (\ref{fact}) above must both be independent of the 
scales $\m$ that define the parton distributions:
$\m d\sigma_{\rm phys}/d\m=0$.  This self-consistency requirement is
readily translated into the ``DGLAP" equation for the evolution
of parton densities,
\be
\mu{d f_{a/h}(\xi,\mu)\over d\mu}=\sum_b P_{ab}(\xi/\eta,\as(\mu))\otimes f_{b/h}(\eta,\mu)\, .
\label{evol}
\ee
Here, the dimensionless kernel $P$ depends only
on variables that are in common between the hard
scattering functions and the parton distributions:
$\alpha_s$ and the momentum fractions.  The scale-independence of
physical quantities and their relations can be studied
systematically \cite{commensurate}.

The idealized pattern for determining and applying 
the distributions may be summarized as follows.
We measure one cross section, $\sigma_{\rm phys}$, at momentum transfer $Q_0$.
Given a ``next-to-leading" order
calculation of  its hard scattering  functions
 $\hat\sigma_a^{\rm (NLO)}$, we determine
NLO parton distributions
$f_{a/h}^{\rm (NLO)}$ at $\m=Q_0$.  Using evolution,
we can then predict  $\sigma_{\rm phys}$ for
any hard process at all $Q$.

The coefficients $c_n$ in Eq.\ (\ref{IRS}) are
known for many processes to NLO \cite{glover}.   They have been
determined at 
NNLO for inclusive DIS and Drell-Yan \cite{nnlo},
and even to three loops \cite{3loop} for selected quantities.
Generally, two loop corrections are available only for single-scale
processes, and the calculation of two-loop corrections for 
genuine scattering diagrams is
an as-yet unsolved, but actively studied, problem in QCD \cite{kaufmann,bern}.

Perturbative QCD is most successful for
inclusive processes, and/or single-scale
semi-inclusive.  Evolution in DIS is perhaps still the best
illustration (see below).  The current experimental
situation in hadronic hard-scattering is reviewed in Ref.\ \cite{huston}.
In multiscale problems, logarithms of ratios of distinct
but perturbative scales 
often require resummation to all orders. 
Formally beyond perturbative resummation, but not always
less important numerically, are corrections suppressed by powers
of the hard scale(s).  In DIS, and a few other cases, these corrections
can be described by the operator product expansion.
We shall encounter below ``generalizations" of this famous
technique, usually in the form of effective field theories. 

\subsection{Prime Examples}

\begin{figure}[h]    
\begin{center}      
\mbox{\epsfysize=8cm \epsfxsize=9cm \epsfbox{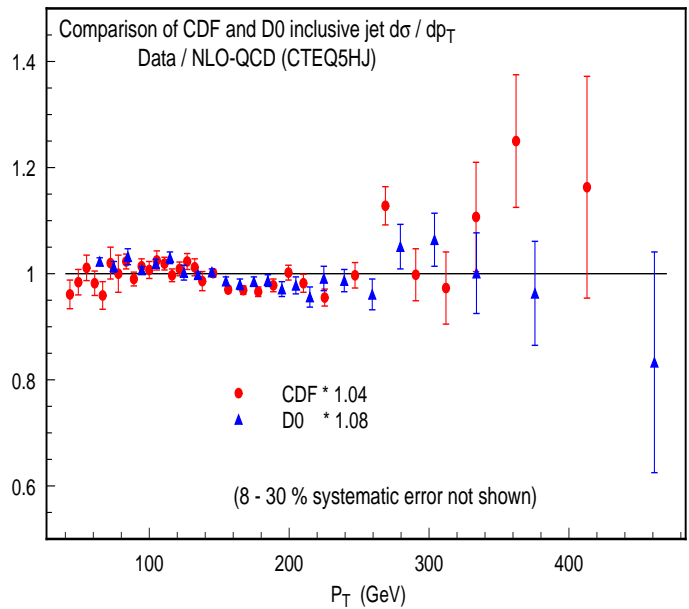}}
\caption{A comparison of single-jet inclusive cross sections
to NLO theory from Ref.\ {\protect\cite{cteq5}}.}
\label{cteqSinglejet}
\end{center}
\end{figure}

{\it Tevatron Jets.} The most impressive success in orders of
magnitude continues to be the Tevatron inclusive
single-jet and dijet cross sections \cite{huston,bhatta,hauser,krane}, as illustrated
in Fig.\ \ref{cteqSinglejet}, which shows a plot from Ref.\ \cite{cteq5}.
According to Eq.\ (\ref{fact}), these cross sections are of the form
\be
\sigma_{p\bar p\rightarrow J+X}=\sum_{ab} f_{a/p}\otimes
\hat \sigma^{\rm (NLO)}_{ab\rightarrow J+X}\otimes f_{b/\bar p}\, .
\label{1jet}
\ee
We find a consistency between NLO theory and experiment at a few tens of percent,
well within the overall systematic errors,
over a range in which the cross section decreases by seven or so factors
of ten.  As the figure shows, reasonable choices of parton distributions
(in this case CTEQ5HJ)
can account even for the highest-momentum data, although a slight difference
remains between the D0 and CDF data at the high end.

\bigskip

{\it DIS Scaling violations.}  Next, we should cite
measurements of DIS structure functions \cite{doyle,erdmann,cross}
for $\ell^\pm N\to \ell^\pm N$, through the
cross sections
\be
{d\sigma^\pm\over dxdQ^2} = {2\pi\alpha^2\over xQ^4}\, 
\left[ \left(1+(1-y)^2\right)F_2(x,Q^2)-y^2F_L(x,Q^2) \mp\left(1-(1-y)^2\right)F_3(x,Q^2)\right]\, ,
\ee
at HERA, the Tevatron, SLAC and elsewhere, with 
$x=Q^2/2p\cdot q \quad q^2=-Q^2 \quad y=Q^2/xs$.
Because the evolution is universal in the factorized forms,
\bea
F_{h}(Q^2) &=& \sum_a C_{a}(Q^2/\mu^2) \otimes f_{a/h}(\mu^2)\; \big |_{Q=\mu}
\nonumber\\
\mu{d f_{a/h}\over d\mu} &=& \sum_b P_{ab}(\as(\mu))\otimes f_{b/h}\, ,
\label{factevol}
\eea
we may think
of evolution as in, but not of, the nucleon, except
perhaps in the  smallest-$x$ region, where target-dependent shadowing,
in which partons begin to interfere with each other, comes into play.

\begin{figure}[h] 
\begin{center}      
\mbox{\epsfysize=8cm \epsfxsize=8cm \epsfbox{14-02fig2a.epsi}}
%\caption{DIS data at moderate to large $x$.}
%\label{H11}
\end{center}
\end{figure}

\begin{figure}[h] 
\begin{center}        
\mbox{\epsfysize=8cm \epsfxsize=8cm \epsfbox{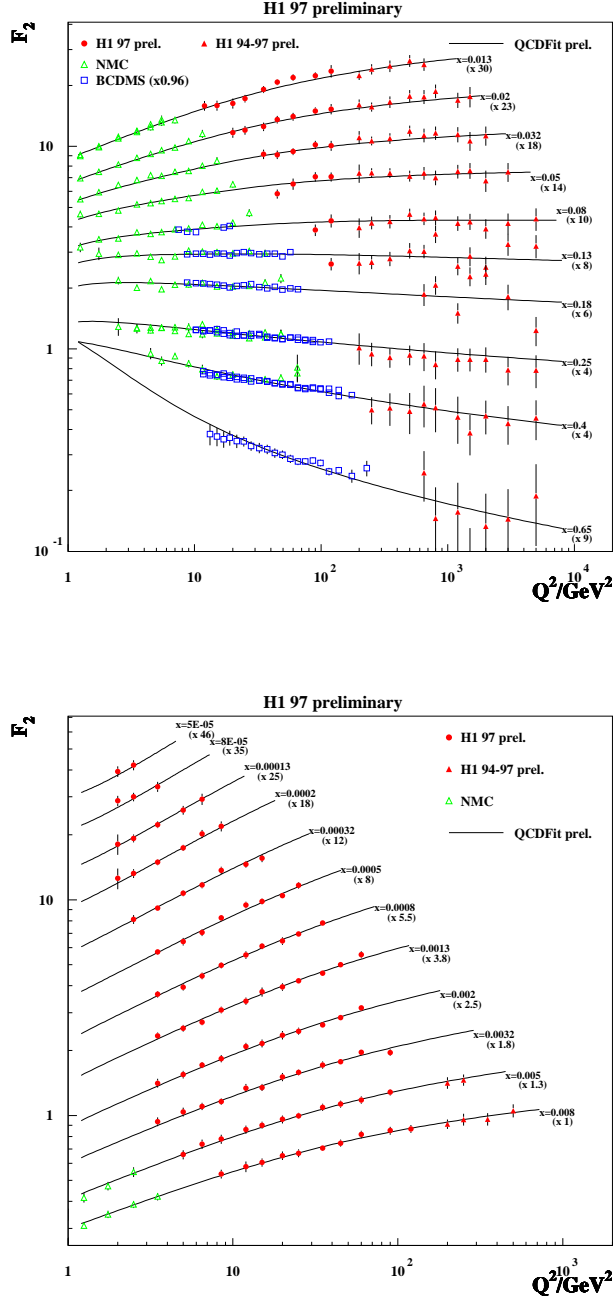}}
\caption{DIS data from H1 (preliminary) {\protect\cite{H1data}}, NMC and BCDMS.}
\label{H12}
\end{center}
\end{figure}

The quality of the data, and of QCD fits to it, is illustrated 
by recent results on $F_2(x,Q^2)$ from H1 \cite{H1data}, shown in Fig.\ {\ref{H12}}
for various fixed $x$ as a function of $Q^2$.  Notice
in particular the sharp rise of $F_2$ with $Q^2$ for the
smaller values of $x$. This is a prediction of the evolution
through parton branching described by Eq.\ (\ref{factevol}).  Deviations
from such DGLAP evolution have been surprisingly
difficult to find.  The interesting results of a 
next-to-next to leading order analysis 
of DIS has been described in Ref.\ \cite{nnlodis}.
At the other end of the spectrum, the excess of  events
reported two years ago  at the largest $x$ and $Q^2$ has
all but disappeared \cite{highq}.

{\it Multijets, and event shapes.} 
In $\e^+\e^-$ annihilation, the Born level process $\e^+\e^-
\rightarrow q\bar q$ produces two jets, as is the case
as well for hadron colliders, while in DIS, the Born
scattering produces a single jet.  In all three
``canonical" scattering processes, the cross sections
for the minimal reactions are known at NLO.
Beyond this, NLO cross sections are available for
two jets in DIS \cite{zeppen}, for  four in $\e^+\e^-$ \cite{dixon}, and
NLO three-jet cross
sections are just now becoming available for hadron-hadron
scattering \cite{kilgore}. Recent years have seen an explosion of data
on these processes, and new results were discussed
at this conference in \cite{bhatta,hauser,krane,zeppen,gary}, along
with jet fragmentation properties \cite{gary,klapp,abe,dong,schyns,lamsa}.
The production of vector bosons associated with jets \cite{croninh}
is another important source of information
about short-distance dynamics, and a reanalysis
of W+1 jet cross sections has brought them back into agreement with theory
over the past year.

Jets are theoretical-phenomenological constructs \cite{tkachov}.
Their value is not that they are an exact reflection of
short-distance reactions, but that, if they are
defined properly, they are related to them in a calculable fashion.
Jets are generally defined in terms of energy flow into
some angular region, or in terms of interative clustering schemes
for the momenta of observed particles.  Any (sufficiently smooth
\cite{irshapes}) quantity
that is insensitive to the emission of zero-momentum
particles, or to the collinear branching of finite-energy massless
particles, can be used to define an infrared-safe cross section.
Event shapes  in $\e^+\e^-$ annihilation
or DIS are chosen for this property.  The best-known is
the thrust, defined for individual events as the maximum fractional projection
of the momenta of observed particles on an axis, as that
axis is varied about the unit sphere.  
Although event shape cross sections are infrared safe, they
generally receive nonperturbative corrections that
decrease rather slowly with energy, typically as
a single power \cite{webberlect},
\be
\sigma_{\rm phys}=\hat \sigma^{\rm PT} \left(1+ {\cal O}(1/Q)\right)\, .
\label{1overQ}
\ee
The theory of these power corrections is a passageway between
the short-distance perturbative, and the long-distance
nonperturbative aspects of QCD.   We shall have more
to say about them below.

{\it The $\as$ lineup.}  Any of the short-distance
cross sections above allows a determination of the strong coupling.
For example, in a factorized cross section, Eq.\ (\ref{fact}),
given parton distributions $f$, we compare $\hat\sigma^{\rm PT}(\as)$ to experiment,
and solve for $\alpha_s$, typically evaluated at a
renormalization scale equal to the factorization scale.
Other, more refined, choices \cite{commensurate} are related by perturbative
corrections.  For cross sections that are infrared
safe, as jet or event shape cross sections at LEP \cite{duchesneau}, the 
comparison is even more direct, although power
corrections, as in Eq.\ (\ref{1overQ}), must be  taken into account for any precise
determination \cite{gary,dokshitzer}.  
Yet another way of determining $\as$ is from
lattice QCD calculations of energy level
differences in heavy quark systems.  These may be
related, on the one hand to the strong coupling, and on
the other to experiment, which then determines the size of the coupling
at a scale that grows with the heavy quark mass.

\begin{figure}[h] 
\begin{center}        
%\vbox{\vskip 7 cm}
\mbox{\epsfysize=7cm \epsfxsize=7cm \epsfbox{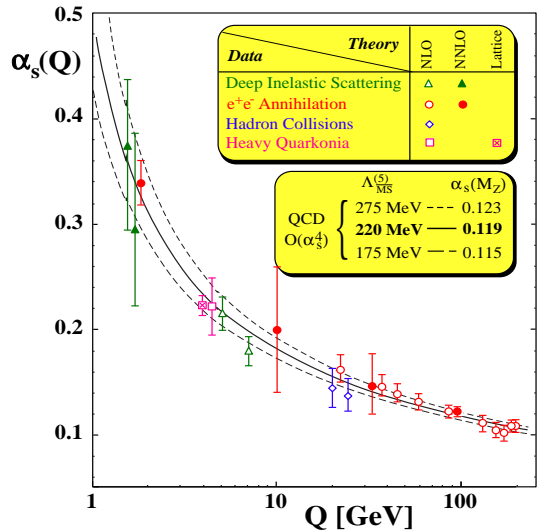}}
\caption{The strong coupling measured at a different length scales,
from Ref.\ {\protect\cite{Betheke98}}.}
\label{alphafig}
\end{center}
\end{figure}

For ease of comparison, the couplings can be evolved to the mass of the Z,
at which a ``world average" of $\alpha_s(M_Z)=0.119 \pm 0.004$ has been quoted
in Ref.\ \cite{Betheke98}.  
Since varied experiments measure the coupling over a wide range of scales, Fig.\ \ref{alphafig},
which shows the variation of measured $\alpha_s(\mu)$ with $\m$ illustrates
one of the great successes of renormalized field theory.

\subsection{Resummation} 

In the presence of multiple scales, high order contributions to
hard scattering functions $\hat \sigma^{\rm PT}$ can become important.
For example, in Eq.\ (\ref{fact}), the limit $x/\xi\to 1$ is
associated with integrable singularities that usually
enhance the cross section.  These ``threshold" singularities were
resummed long ago for the Drell-Yan cross section, and have
been applied for some time
to top production at leading-log accuracy \cite{topresum}. In QCD hard-scatterings, 
such as heavy quark and jet production, color exchange makes the resummation
somewhat more complex beyond leading log.  This problem has now been solved,
and threshold resummation is understood in principle
for a wide variety of hard scattering 
cross sections \cite{colorresum1,colorresum2,colorresum3}.  As a practical
matter, differing approaches to the inversion of certain Mellin transforms 
can lead to differing numerical predictions.  From a broad
perspective, however, the main lesson is that the theory stays
relatively close to NLO for cross sections like top production, even
in the presence of superficially large corrections at higher order.
An important observation \cite{topresum,colorresum2}, made particularly
clear in the very recent publication \cite{colorresum3}, is a marked
decrease in  factorization-scale dependence for resummed
cross sections.

\begin{figure}[h]    
\begin{center}        
\mbox{\epsfysize=8cm \epsfxsize=8cm \epsfbox{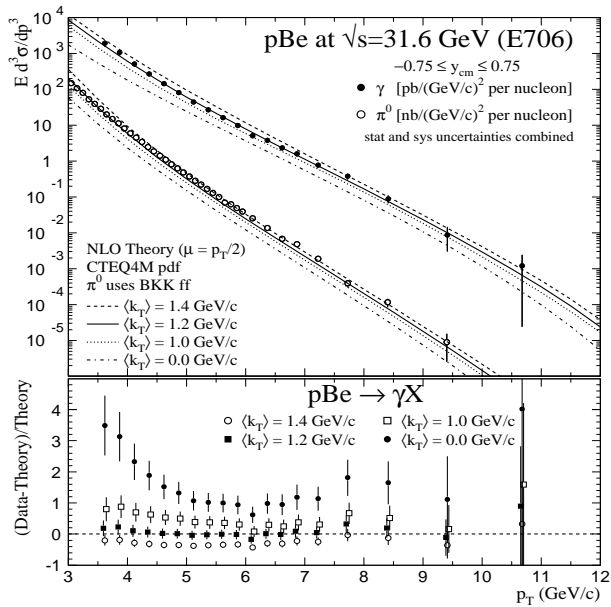}}
%\epsffile{E706plot.epsi}
\caption{E706 data versus NLO predictions and $k_T$-smearing.}
\label{E706fig}
\end{center}
\end{figure}

Another example, currently being discussed widely,
is related to the data of Fig.\ \ref{E706fig}, single-photon and pion
inclusive cross sections 
measured by the fixed-target Fermilab experiment E706 \cite{begel,e706}. 
So far, this data can be fit only by 
supplementing NLO with the old method of $k_T$ 
smearing for the initial partons \cite{706cteq}.
At the same time, it
has been noted that the full range of fixed-target direct photon data
may not be consistent among themselves \cite{aurenche}.
It should also be noted that
at collider energies, experiment and NLO agree, at least
at the higher transverse momenta \cite{gordon}.

Can the resummation of higher orders in QCD lead to an effective 
$k_T$ smearing?  For W or  Z production at low transverse
momentum, the answer is yes, and the formalism has existed for 
some time.  In this case, fitting the cross section \cite{d0wpt}
requires the introduction of nonperturbative parameters that
are accessible to experiment.  In the past year or  two, there
has been some discussion on the best way of going about this
 \cite{ellisvesl,yuanmor}, but the 
underlying theory is relatively well-understood.
The same cannot be said for W or photon cross sections at
high transverse momenta, because the logarithms
whose resummation requires the nonperturbative input at $Q_T\sim 0$
cancel order-by-order in perturbation theory in the
calculation of $\hat \sigma$ at high $Q_T$.  Interesting
observations on the relationship between the two regimes
have been made, however \cite{likt}, and futher progress in this
direction can be anticipated in the coming year.

One of the major challenges in perturbative QCD is the development
of a theory of these and related higher-order effects, and a method for estimating their
importance.  We shall return to developments on
resummation below.

\section{The Long and Short of Hadron Structure}

\subsection{Parton Distributions 1999}

The parton distribution functions (PDFs) in Eq.\ (\ref{fact}) summarize the
structure of hadrons as seen at short distances, one parton
at a time.  As the hadron is probed at ever shorter distances,
the distributions evolve perturbatively according to Eq.\ (\ref{evol}), 
but the boundary condition for this evolution stands as
a truely nonperturbative reflection of the dynamics that
holds the hadron together.  These parton distributions have 
been the subject of intense study since the recognition of
approximate scaling in DIS structure functions thirty years ago.

Over the past decade, two groups, CTEQ and MRS, whose memberships
have themselves evolved somewhat, have undertaken coordinated
``global" approaches to the determination of parton distributions,
taking into account data from a variety of processes and momentum
scales.  The past year has seen the development of the
latest, best fits of these two groups, MRST \cite{mrst}
and CTEQ5 \cite{cteq5}, which were discussed and compared
at the conference in \cite{wkttalk}.  

Global fits test the self-consistency of factorized cross
sections in the sense that the fits are overconstrained,
and because they can be checked against experiments not 
incorporated into the fits.  Nevertheless, the fits must be
improved as the data improves, and as it extends to extreme
values of fractional momenta.  Surprises can occur, especially
in  regions where a particular parton distribution does
not contribute at leading order.
Examples of such refinements from 1998 involved the
ratio of $d$ to $u$ quarks in the proton, as
tested by the W asymmetry \cite{Wasy} and Drell-Yan \cite{DYdtou}.
Generally speaking, indirect constraints on
parton distribution functions are tentative.

Other cases where the results of global fits have been
rethought involve 
higher-order or power (``higher-twist") corrections
to the cross section.
The extraction of PDFs 
requires a parameterization of such effects, assuming that
they can be brought under control. 
Examples include the role of higher twist in the
extraction of neutron PDFs from deuterium data \cite{yangbodek}.

Up to this cycle of global fits, the primary processes employed
were DIS, Drell-Yan and direct photons, the latter thought
to be especially valuable for constraining the gluon 
distribution, which does not appear at leading order in
the other two.  The data of E706 \cite{begel,e706}, however,
a sample of which was shown above in Fig.\ \ref{E706fig},
has thrown this neat picture into disarray, as
it disagrees decisively with the predictions of Eq.\ (\ref{fact})
at NLO.  A $k_T$ smearing approach 
has been used in the MRST fits of last
year \cite{mrst}.  The CTEQ5 fits abandon  direct
photons in favor of jet cross sections to constrain the
gluon at large $x$ \cite{cteq5}.  

Yet another interesting problem is the  treatment of heavy quarks.  For energies
near a heavy-quark mass, $m_Q$, their production may be
calculated directly as a hard process, part of $\hat \sigma^{\rm PT}$
in Eq.\ (\ref{fact}).  For energies much above $m_Q$,
however, it may be advantageous to treat the heavy
quark as a parton,
thus automatically resumming logs of $m_Q^2/s$.
A number of schemes to make this transition have been
proposed \cite{cteq5,mrst,svnhq,thornerv}, to treat 
charmed quark production at HERA.

The sophistication of these considerations of global analysis,
and the need to make accurate predictions at high energy
raises the question of
 how to estimate uncertainties in the distributions \cite{ctequncer}.  
In part to explore these issues, new sets of distributions
have been produced based on DIS data only \cite{disPDF},
and methods have been introduced to quantify uncertainties
systematically through statistical analysis \cite{bayseanpdfs}.

\subsection{Spin and Off-diagonal Distributions}

The past few years has seen a rebirth of interest in 
the high-energy physics of spin, which, with
advances in the technology of polarized beams and targets,
has made possible the systematic study of spin at the
parton level \cite{doyle}. 
The DIS cross section for a nucleon with spin $s$ may be represented in
the standard form
${d^2\sigma/d\Omega dE}
=
(\alpha_{\rm EM}^2/2mQ^4)(E_e/ E'_e)
L^{\mu\nu}W_{\mu\nu}$, in terms of
spin structure functions defined as
\be
W_{\mu\nu} = W_{\mu\nu}^{\rm unpol} + {i\over E_e-E'_e}\epsilon_{\mu\nu\lambda\sigma}q^\lambda s^\sigma\; g_1(x,Q)
 + {i\over (E_e-E'_e)^2}\epsilon_{\mu\nu\lambda\sigma}q^\lambda\left[p\cdot q s^\sigma
- s\cdot q p^\sigma\right ] g_2(x,Q)\, .
\label{spinstructure}
\ee
The function $g_1$ has a particularly transparent interpretation
in terms of quark helicity:
$g_1(x,Q) = {1\over 2}\sum_f e_f^2\Delta  q_f(x,Q) + O(\alpha_s)$,
where $\Delta q_f$ is the  difference between the distributions
of quarks with helicity parallel to the hadron's helicity and against it.
At this conference, precision data for $g_1^p$ and $g_1^n$
were presented by E155 \cite{mitchell}.

\begin{figure}%[t]    
\begin{center}        
\mbox{\epsfysize=9cm \epsfxsize=9cm \epsfbox{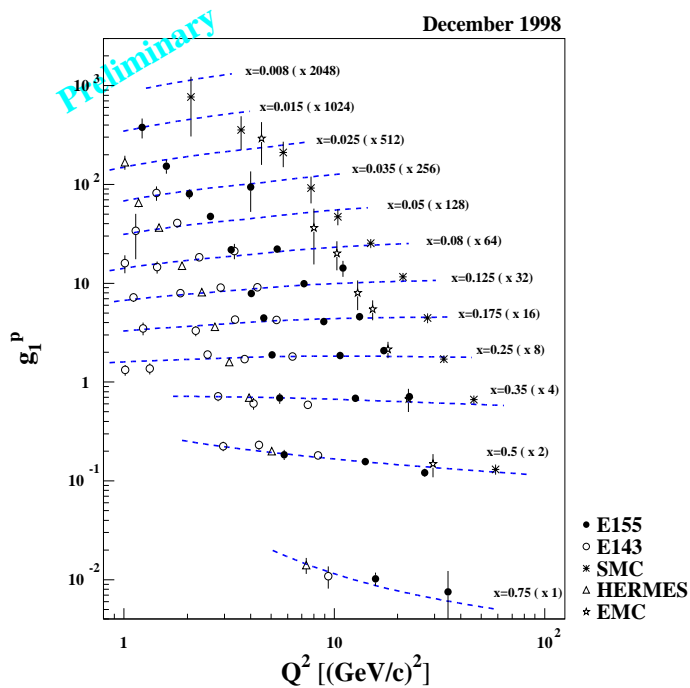}}
\caption{E155 measurement of $g_1$ {\protect \cite{mitchell}}.}
\label{E155fig}
\end{center}
\end{figure}

One result of these measurements
is a test of of the benchmark Bjorken sum rule,
\be
\int_0^1 \left( g_1^p(x)-g_1^n(x)\right)dx = 
{1\over 6}\; \left|{g_A\over g_V}\right|\; \left(1+\sum c_n\alpha_s^n\right)\, ,
\label{bjsr}
\ee 
which is now verified to a new level of accuracy.

The ``spin" distributions of the nucleon, such as $\Delta q_f$, do
not necessarily describe its full spin content, and the possibility of orbital angular
momentum must also be taken into account.  At the
same time, for gluons the distinction between these two types
of angular momentum is not gauge invariant.  
This problem notwithstanding, an attractive formalism for
the description of orbital angular  momentum has been proposed
\cite{xji}, in terms of form factors,
$J_{q,g} = {1\over 2}\; \left[A_{q,g}(0)+B_{q,g}(0)\right]$,
which arise in matrix elements of
the energy-momentum tensor,
\be 
\langle p+\Delta|\; T_{q,g}^{\mu\nu}\; |p\rangle
= \bar u(p+\Delta)\Big[\; A_{q,g}(\Delta^2)\; \gamma^{\mu}\bar p{}^{\nu}
+B_{q,g}(\Delta^2)\; \bar p{}^\mu{i\sigma^{\nu\alpha}\over 2M}\Delta_\alpha
-(\mu\leftrightarrow \nu)\; \Big]u(p)\, ,
\label{offdiagonal}
\ee
with $\bar p=p+\Delta/2$.
At zero momentum transfer, the $J_{q,g}$ become expectation values
of the angular momentum operators
\be
{\bf J}_q = 
\int d^3x\; \left[\; \psi^\dagger {{\bf \Sigma}\over 2}\psi+\psi^\dagger {\bf x}\times(-i{\bf D})\psi\; \right]\, ,
\quad 
{\bf J}_g =
\int d^3x\; {\bf x}\times({\bf E}\times{\bf B})\, .
\ee
The measurement of these off-diagonal matrix elements, unlike the
diagonal matrix elements that define classic PDFs (see
Eq.\ (\ref{msdist}) below), require the
measurement of exclusive, or semi-exclusive amplitudes,
such as ``deeply-virtual Compton scattering",
$\gamma^*(Q^2)+N(p)\rightarrow \gamma+N(p')$,
with $p^2=p'{}^2=m_N^2$.  The measurement of such amplitudes is a 
challenge, but one that is of great interest for the Jefferson 
Laboratory facility, and there is a correspondingly vigorous
theoretical program to study the factorization
and evolution properties of off-diagonal distributions \cite{radyod,collinsod,balitskyod}.
Off-diagonal matrix elements interpolate, in some ways, between
inclusive and exclusive processes, and between parton distributions
and hadronic light-cone wave functions \cite{brodsky}.  

The consideration of orbital angular momentum leads us to the
doorstep of the long-distance, low-energy properties of hadrons,
where progress has continued in lattice QCD and
instanton studies.

\subsection{Lattice Hadron Spectra, Quark Masses}

Lattice methods \cite{sharpe} approach QCD from a limit complementary to perturbation
theory, and make possible direct calculation of long-distance properties of hadrons.
Typically, lattice computations involve the 
evaluation of expectations of nonlocal products of operators, such as
\be
C_J = \langle 0\; |J(x)\; J(0)|\; 0\rangle\, ,
\quad J = \prod_i \bar q_i(x+\Delta)\; O_{ij}\; q_j(x-\Delta)\, ,
\label{lattvev}
\ee
with the $q_i$ quark fields,
where $O_{ij}$ projects a set of quantum numbers.
For $x\rightarrow \infty$,
the $x$-dependence is dominated by the energy of the lowest-lying state(s)
of the relevant quantum numbers.
As noted above, such studies can be  used to set the scale for
$\alpha_s$ by studying the hyperfine splitting in heavy quarkonia.

The numerical evaluations of expectations like (\ref{lattvev}) are said to be  quenched,
partially quenched, or fully unquenched, depending on how many species light fermions are
allowed to fluctuate out of the vacuum.  At the conference, 
progress was reported in quenched and unquenched lattice QCD,
on the calculation of realistic spectra for hadrons, and of
matrix elements of hadrons involving both heavy and light quarks \cite{kuramashi,gottlieb}.
Variations on this theme are now making possible the calculation of 
realistic decay matrix elements \cite{gottlieb,simone,pukurovsky}.
In addition, they allow the exploration of ideas on the
mechanisms of confinement \cite{cornwall}.  In alternative formulations,
lattice methods may be applied to light-cone formulations of
hadronic structure \cite{brodsky,vandesande}.

The current sophistication of lattice techniques, and the growing
power of new machines, including Teraflop computers at the
  U.\ of Tsukuba and at Brookhaven(RIKEN)-Columbia, are making possible
a new generation of investigations of chiral symmetry breaking.

Roughly speaking, chiral symmetry in QCD (also discussed at the conference in
\cite{tandean}) reflects the 
observation that
gluon emission doesn't change helicities.  
As a result, in the absence of  quark masses in the propagator, left- and
right-handed quarks decouple altogether in perturbation theory.
This means that a massless
quark stays massless in perturbation theory.  
Chiral symmetry, however, is broken nonperturbatively at zero temperature, but
restored at finite termperatures, a transition which
can also be studied on the lattice \cite{negele,vranas}.
 At the same time, current
algebra requires that the square of the pion mass would
vanish linearly with the lightest quark masses,
\be
m_\pi^2 \sim m_{\rm light}\, .
\label{pionqkmass}
\ee
Historically, this kind of relation has been difficult to realize on the lattice
because of corrections that vanish
as a low power of the lattice spacing in the continuum  limit.  

With the advent of ever more powerful machines, the method of
``domain wall fermions" \cite{kaplan} makes possible new approaches to  
relations like Eq.\ (\ref{pionqkmass}).
In the domain wall technique, the
chiral symmetry is manifest, up to {\it exponential} corrections,
at the price of introducing a  5th dimension, which we label by coordinate $s$,
  in the five-dimensional Dirac Lagrange density
\be
{\cal L} = \bar\psi\; [\gamma\cdot D[A] + m(s)]\; \psi\, ,
\label{wallL}
\ee
where the mass parameter depends on $s$, and vanishes at endpoints 0 and $s_0$:
$m(0) = m(s_0)=0$.
The light quarks are zero modes that propogate in 
``our world" at $s=0$ and $s_0$, one wall for each helicity.
These new methods have shown early success \cite{vranas,blum} in preserving chiral 
symmetry, in terms of relations such as Eq.\ (\ref{pionqkmass}).
In the lattice domain wall construction,  the extra dimension
is just a convenience, but it cannot be denied that
the technique bears an eerie resemblance to the brane constructions
of modern string theory \cite{terning}.

\subsection{Hadrons and Instantons}

Successes in the treatment of chiral symmetry in lattice QCD
lead us naturally to the reemergence of instanton studies
of hadron dynamics \cite{Shuryak}.  
In the language of spontaneous symmetry breaking,  the masses of hadrons 
are related to the generation of quark ``condensates" $\langle \bar q q\rangle
\sim \langle \bar q_R q_L\rangle+\langle \bar q_L q_R\rangle$
in the QCD vacuum, which couple (the perturbatively decoupled)
left- and right-handed components of the quark field.

Instantons may be thought of as tunneling events between the inequivalent QCD vacuum
configurations that are distinuished by different phases of the nonabelian
gauge fields at infinity (even at zero energy).
The instantons couple to the quarks, and the process of tunneling
produces an effective $2N_f$-point interaction between their
left- and right-handed components, of the general form
\be
{\cal L}=G\; {1\over 8N^2}\left[\left(\bar \psi\tau^-\psi\right)^2
+\left(\bar \psi\tau^+\gamma_5\; \psi\; \right)^2\; \right]\, ,
\label{fourquark}
\ee
even though the quarks are massless.

In the ``instanton liquid" model, the nonperturbative soft gluon field
is replaced by an
ensemble of instantons and anti-instantons that couple left- and right-handed 
quark-antiquark pairs.  Meson masses are produced
as pairs ``hop" between instantons, changing helicities as they go.
In this model, mesons and baryons, with realistic spectra and matrix elements
emerge, and provide a cross-check for lattice calculations \cite{negele}
at zero and at high temperatures.

\section{Factorizations, Evolutions and Effective Theories}

Every hard-scattering experiment includes a complete evolution
all the way from short distance to long distance dynamics.  
Factorization allows us to organize 
the long distance dynamics, and thus to calculate perturbative short-distance
dependence, and compare the results to experiment. 
The essence of factorization is to interpret
long-distance information in terms of matrix elements in
the underlying theory.
For example, in the classic factorization in Eq.\ (\ref{factevol})
 for DIS structure functions of hadron $h$,
$F^{(h)}(Q)= \sum_a C_a(Q/\mu)\otimes f_{a/h}(\mu)$,
the quark ($a=q$) distributions may be interpreted as
\be
f_{q/h}(x,\mu)
=
\int_{-\infty}^\infty {dy\over 4\pi}\, \e^{-ixP\cdot ny}\, 
\langle h(P)|\; \bar q(yn)n\cdot \gamma\, \Phi_n(y,0) q(0)\; |h(P)\rangle\, ,
\label{msdist}
\ee
where $n$ is a lightlike vector not in the direction of the
hadron momentum $P$, and $\m$ enters
as the renormalization scale for the matrix element,
which is ultraviolet divergent for $n^2=0$.  The function
$\Phi_n(y,0)$ is a path-ordered exponential of the gauge field,
\be
\Phi_n(y,0)= P\exp \left[ -ig\int_0^y
 d\lambda\; n\cdot A(\lambda n)\right]\, ,
\label{oexp}
\ee
which makes the matrix element gauge invariant. 
The evolution of Eq.\  (\ref{evol}) may be thought of as 
a consequence of the renormalization properties of
the nonlocal operators in $f_{a/h}$, which summarize an
infinite set of twist-two matrix elements in the light-cone expansion.
The same formalism is at the basis of  heavy
quark effective theory and of nonrelativistic QCD (NRQCD) \cite{benekerv}, although
generally with a finite sum over operators rather than a convolution.
In a sense, the operator $\Phi_n$ in Eq.\ (\ref{msdist})
plays the role of the heavy quark field in heavy quark
effective theory, as a nonrecoiling source of gluons. 

Over the past few years, factorization,
analyzed in terms of effective operators, has been applied
to multiscale hard scattering processes, with Sudakov \cite{cls,sotir}
and Regge high-energy limits.  
The former refers to cross sections with large
momentum transfer and low 
QCD radiation, the latter to low momentum transfer
and essentially unlimited radiation.  The Regge limit
is related to the total  cross section, while the
Sudakov limit highlights its short-distance components. 
 
One of the simplest examples of the 
Sudakov limit is the dijet cross section in $\e^+\e^-$ 
at fixed jet masses $m_i^2$, $i=1,2$.  In the limit of 
light jets, $m_i^2\ll Q^2$, the dijet cross section is related
to the integrated thrust cross section by
\be
\int_T^1 dT'\; {d\sigma(T')\over dT'}
= 
\int dm_1^2dm_2^2\; {d\sigma\over dm_1^2dm_2^2}\;
\theta\left( 1-T-{m_1^2+m_2^2\over Q^2}\right)\, .
\ee
The cross section for
``nearly-lightlike" jets in $\e^+\e^-$ satisfies a
factorization \cite{cls}
\be
{d\sigma\over dm_1^2dm_2^2}= {C(Q/\mu,\beta_i)\over Q^6}
\; \otimes\; \prod_{i=1}^2 J_i(\mu,\beta_i) \otimes S(\mu,\beta_i)\, ,
\label{dijetfact}
\ee
up to corrections suppressed by powers of $m_i^2/Q^2$,
with $\beta_i$ the 4-velocity of jet $i$.  In (\ref{dijetfact}),
there is a double factorization, separating the dynamics of
the jets, included in the functions $J_i$, from both the
truely short-distance ``coefficient" function $C$ and from
the dynamics of relatively low-energy  partons 
emitted coherently by the jets and included in 
the function $S$.  The ``soft" function $S$ is 
associated with a particularly interesting composite operator in QCD.
$S$ describes the emission of gluons whose wavelenths are so
long that they cannot resolve the internal structure of 
the jets, and are thus generated by the product
\be
W(0)=\Phi_{\beta_1}(\infty,0)\,
\Phi^\dagger_{\beta_2}(\infty,0)\, ,
\label{Wdef}
\ee
where, in the notation
of Eq.\ (\ref{oexp}), the $\Phi$'s
are ordered exponentials, and
$\beta_1$ is the velocity of the quark jet, and $\beta_2$
of the antiquark jet.  

We need
not dwell on the nature of the convolutions denoted by $\otimes$
in Eq.\ (\ref{dijetfact}), but the 
double factorization itself is adequate to imply
a resummation \cite{cls}
of double logarithms in $1-T$ \cite{cttw},
\be
{1\over \sigma_{\rm tot}}\int_T^1\; dT'{d\sigma(T')\over dT'}
=
\exp\; 
\left[-2C_F \int_{1-T}^1 {d\a\over\a} \int_{\a^2Q^2}^{\a Q^2} 
{dk_T^2\over k_T^2}\; {\as(k_T^2)\over \pi}+\dots\right]\, ,
\label{thrustresum}
\ee
where corrections include fewer logarithms of $1-T$
in the exponent.
These results are also related to the renormalization
properties of the operators $W$ in Eq.\ (\ref{Wdef}).    

Another application of factorization and effective
operators is to resummation in the ``Regge" limit
$s\to \infty$, $t$ fixed, the ``BFKL" regime for QCD.  The BFKL
equation may be derived from a 
 ``multiperipheral" reexpression of DIS factorization,
Eq.\ (\ref{factevol}) \cite{tasi95,balitsky,kw_lievol}:
\bea
F(x,Q^2)&=& \int_x^1 {d\xi\over\xi}\; 
C\bigg({x\over\xi},{\mu^2\over Q^2}\bigg)G(\xi,Q^2)+{\cal O}\left ({1/Q^2}\right )
\nonumber \\
&=& \int d^2k_T\; c\bigg({x\over\xi'},Q,k_T\bigg)\psi(\xi',k_T)\
+{\cal O}\left({1/\ln (1/x)}\right )\, ,
\label{xrefact}
\eea
where the $k_T$-dependent distribution $\psi$ is
related to the gluon PDF by \cite{ccfm}:
\be
G(\xi,Q^2)=\int^Qd^2k_T\; \psi(\xi,k_T)\, .
\label{psiktdef}
\ee
In the second form of Eq.\ (\ref{xrefact}), the roles of
longitudinal and transverse momenta have been reversed, and
corrections are suppressed only by logarithmics of $x$, rather than
powers of $Q$.  In these terms, the BFKL equation,
\be
\xi{d\psi(\xi,k_T)\over d\xi} = \int d^2k'_T\; {\cal K}(k_T,k'_T)\psi(\xi,k'_T)
= -{\as N\over \pi^2}\, \int {d^2k_T\over (k_T-k_T')^2}\; 
\left[\; \tilde \psi(\xi,k_T) - {k_T'{}^2\over 2k_T^2}\tilde \psi(\xi,k_T')\; \right]\, ,
\label{bfkl}
\ee
with $\tilde\psi\equiv(1/k_T^2)\psi$,
describes the evolution of $\psi(\xi,k_T)$ in $\xi$.  The same
equation may also be derived from the renormalization of ordered exponentials,
like Eq.\ (\ref{oexp}).  Indeed, distributions of exponentials integrated
over a transverse density \cite{balitsky}
\be
\int d^2x_T\; \rho(x_T)\; P\; \exp\left[\; -ig\int dx^+A^-(x^-,x_T)\; \right] 
\label{distribution}
\ee
are being studied to develop
``unified"  effective theories that describe
the variety of evolution equations, including
DGLAP, BFKL and others \cite{kw_lievol}.

\section{BFKL, Diffraction and Color Dynamics}

\subsection{BFKL at NLO}

What's so special about the BFKL equation?  It addresses
the total cross section in gauge theory in terms of its
fundamental quanta, the quarks and gluons.  Now it is
not entirely obvious that such a project will work, ultimately,
in an asymptotically free theory with confinement,
but if it does, it will say something fundamental about field theory.
In an older language, this would be a theory of the ``pomeron" \cite{levin1}. 
Also, in the language of the parton
model, BFKL appears to predict that, as 
we evolve to low $x\ \leftrightarrow$ high $s_{\gamma^*N}$ at fixed $Q$, we reach a
region of high parton density
at nearly fixed (actually slowly diffusing \cite{muellerdiff}) virtuality.  This
is a new ``intermediate" regime of QCD, between perturbative
and hadronic phases \cite{mueller98}.  It is relevant to the
screened ``plasma" state, which we hope to encounter at RHIC.

The solutions to the lowest order BFKL equation are of the form
\be
\tilde \psi \sim x^{-\o}\; \left({k_T^2\over\mu^2}\right)^{-i\nu-1/2}\, .
\ee
The largest permissible value of $\omega$ gives the
dominant low-$x$ behavior, which is found to be
\be
\nu=0, \quad \o= \o_0 \equiv 4N\ln2\;(\as/\pi)\, .
\label{losoln}
\ee
From the kinematic relation in DIS, $s_{\gamma^* N}= Q^2(1-x)/x$,
the low-$x$ behavior of $\psi(x,k_T)$ determines the large-$s$
behavior of the $\gamma^* N$ total cross section, for
which the lowest-order BFKL result (\ref{losoln}) gives
\be
\psi \sim ``\sigma_{\rm tot}"\sim s^{4N\ln2\;(\as/\pi)}\, .
\label{sigtot}
\ee
This is a derivation of Regge-like behavior 
for the total cross section from perturbative QCD,
and, because it involves the exchange of no overall
quantum numbers, may be considered as a perturbative model
for the pomeron.

\begin{figure}[ht]
\centerline{\epsfysize=6cm\epsfxsize=6cm
\epsffile{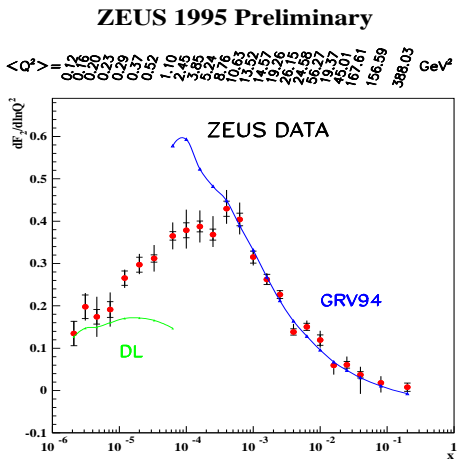}}
\caption{Slopes of $F_2$ as measured by
ZEUS.}
\label{slopeplot}
\end{figure}

Where should one look for BFKL behavior in experiment?
Suggestions include correlations in dijet and rapidity-gap
cross sections in DIS and ${\rm p}\bar{\rm p}$,
and in non-DGLAP evolution in DIS at low $x$ and moderate $Q^2$.
In the first case, there may be hints in the 
dijet data from HERA and in the comparison of 
jet correlations at 630 and 1800 GeV \cite{krane}.
In the later case, the strong rise in $x$ of the structure
functions $F_i(x,Q^2)$ cannot be sustained indefinitely, since,
by (\ref{sigtot}), this would eventually violate unitarity bounds
for the cross section.  Before this happens, interference
between partons, or ``shadowing",
which is absent in both DGLAP and BFKL evolutions,
 must begin to set in \cite{gotsman}.  One  of the 
much-discussed data presentations
of the past year, Fig.\ \ref{slopeplot} from the ZEUS collaboration, 
shows the transition between perturbative and nonperturbative
behavior in a particularly suggestive form \cite{zeusv}.

1998 was the year of the NLO BFKL kernel, the year in which the decade-long
project of computing the next-to-leading order kernel in Eq.\ (\ref{bfkl})
bore fruit \cite{nlo2loop}.  At NLO, $\cal  K$ is fairly complicated,
but the effect on $\o$ in Eq.\ (\ref{losoln}) is simple  enough,
\be
\o^{(NLO)}= \o_0\; \left[1-6.6\left({N\as\over\pi}\right)\right]\, .
\label{omeganlo}
\ee
Now 6 and a half is not by itself a large number, but
the size of this correction nevertheless presents a challenge, because unless
$\alpha_s$ is quite small, the second term may overwhelm the first and
lead to an unrealistic falling cross section.  In addition, it
has been observed that the NLO kernel even implies a non-Regge behavior
at high orders \cite{nlorun}.  Interesting responses to these challenges were
discussed in \cite{schmidt}.  Evidently, the NLO fruit of
BFKL will be an acquired taste, but the coming year 
surely promises intensive work and further clarification

\subsection{Diffraction and Diffractive PDFs}

Through the optical theorem, the total cross section is closely
related to elastic and diffractive scattering amplitudes \cite{erdmann}.  
In  diffractive DIS, we  can relax inclusivity and
thus probe QCD dynamics in the final state, while retaining a large momentum
transfer.  Convenient variables to describe diffraction are
\be
x_{\cal P}={M_X^2+Q^2\over W^2+Q^2}\ \quad \beta={x\over x_{\cal P}}\, ,
\label{diffx}
\ee
where $M_X$ is the mass of an observed system $X$ moving
in the ``current" ({\it i.e.} photon)
direction,
and $W^2=s_{\gamma^*s}$.  $x_{\cal P}$
is the fractional longitudinal momentum transfer from the nucleon to
$X$, and $\beta$ is the equivalent fractional momentum of
a parton in the (hypothetical) exchanged ``pomeron".

Diffractive events are typically defined by
a large gap in rapidity between $X$ and the elastically-scattered, or diffractively-excited,
proton (or its low-mass fragments), which has experienced invariant momentum
transfer $t$.
In these terms, a fully differential cross section is
\be
{d^4\sigma \over dQ^2 d\beta dx_{\cal P} dt} = 
{2\pi\alpha_{\rm EM}^2\over \beta Q^4}\; \left( 1+(1-y)^2\right)\;
F^{D(4)}_2\, .
\label{diffcrossx}
\ee
In the simplest diffractive processes $X$ consists of a single
vector boson.  In this case, the relevant amplitude is
factorized in terms of
off-diagonal PDF's, related for small $x$ to the
gluon distribution, which behaves as $G(x)\sim x^{-\lambda}$.
The resulting cross section is proportional to $G^2(x)$, and hence 
increases with $W$, as
\be
{d\sigma^D\over dM_X^2}
\sim
{1\over M_X^4}\; W^{4\lambda(M_X)}\, .
\ee
There is compelling evidence for this
behavior, with a value of $\lambda$ increasing with $M_X$,
suggesting once again a Regge-like behavior reminiscent
of BFKL.  This connection has yet to be completely
explored.

High-$Q^2$ DIS diffractive cross sections
may be factored \cite{difffactprf} using specifically diffractive PDFs, 
also referred to as fracture functions \cite{fracture},
\be
F^D=\sum_a\; C_a^D\otimes f_a^D\, .
\label{diffact}
\ee
Phenomenological
fits to $f_{q,g}^D$'s have been carried out \cite{diffpdf},
and lead to  predictions whenever a factorization 
like Eq.\ (\ref{diffact}) applies \cite{hautmann}.  It is important
to realize, however, that because diffractive PDFs are
not fully inclusive, they depend on the details of evolution into
the final state, in particular the likelyhood of the
target proton staying together.  This probability cannot
be expected to be the same in ${\rm p}\bar{\rm p}$
cross sections, where the fragments of two initial-state
hadrons pass through each other, as in DIS, where only
a single hadron is involved. And indeed, studies \cite{alves} have
shown that diffraction at the Tevatron is much less
likely than would be suggested by a direct generalization
of Eq.\ (\ref{diffact}) to this case with universal
diffractive PDFs.

Nevertheless, double-diffraction (double rapidity gap) jet production is
seen at the Tevatron, and the jets  show a standard parton-parton $E_T$-dependence
\cite{alves},
indicating that the short-distance process is independent
of the long-distance evolution.  This suggests that 
another factorization is possible in this case, and may
shed light on diffractive dynamics.

\subsection{Color Dynamics}

Diffractive processes are naturally interpreted in terms of color-singlet
exchange in the $t$-channel.  Large momentum transfer processes, however,
are described perturbatively in terms of single-gluon exchange, carrying
octet quantum numbers.  Over the past few years, there has been 
progress in understanding the relation between these two pictures.

Although color is not observable, representations ($1,q=3,g=8\dots$) are,
at least in principle.  In NRQCD, a factorization
that includes the mixing of operators with differing color content has
already led to valuable insights and phenomenological successes \cite{benekerv}.
For high-energy processes, rapidity gaps were long ago
suggested by Bjorken as an ideal arena to study color exchange,
with singlet exchange expected to produce an excess of events with
very low interjet multiplicity.  

It is clear, however, that it is not possible to separate
short- from long-distance color exchange uniquely, since gluons of all
momenta carry the same color content, and models in which 
the color content of the final state is determined at the longest
distances have had success \cite{halzenzepp}.  At the same time, energy flow into
regions between two high-$p_T$ jets is senstive to all time scales
between $1/|p_T|$ and $1/\Lambda_{\rm QCD}$.  Energy flow at the
shorter time scales is both perturbative and sensitive to the
color content in the $t$-channel.  This observation led \cite{OdSt}
to an analysis of dijet cross sections in terms of energy flow $Q_c$ into the
interjet region, in the range $\Lambda_{\rm QCD}\ll Q_c\ll \sqrt{-t}$.  
The cross section at fixed $Q_c$ may be
factorized, and logarithms of $Q_c^2/t$ resummed.  
This behavior is found from the renormalization properties
of composite operators that are products of ordered exponentials,
which generalize
Eq.\ (\ref{Wdef}) to the $2\to 2$ scattering of partons with
color exchange (labelled here by f),
\be
W^{\; \rm f}_{\beta_j\alpha_j}(0) = \prod_{i=3}^4 \Phi_{n_i}(\infty,0)_{\beta_i,\kappa_i}
T^{\; \rm f}_{\kappa_4\dots \kappa_1} \prod_{i=1}^2\Phi_{n_i}(0,-\infty)_{\kappa_i,\alpha_i}
\label{Wfdef}
\ee
where $T^{\; \rm f}$ is a matrix that couples the color of the incoming
and outgoing ordered exponentials $\Phi_{n_i}$, representing the active
partons of the hard scattering.  These operators mix under renormalization
and induce an evolution that tracks
 mixing in the color space as the scattering particles
(including $q\bar q,\, qg,\, gg$) evolve from short to long distances.
This analysis offers a new set of predictions for $p_T$, energy and
rapidity dependence of gap events, which can be tested at Run II of
the Tevatron, and at the LHC.

\section{Power corrections}

There has been considerable interest in power
corrections to infrared safe quantities.  As noted in
Sec.\ 2 above, such power
corrections are quite important in the phenomenology
of jet cross sections and event shapes in $\e^+\e^-$ annihilation \cite{gary,duchesneau}.
Behind this work is a hypothesis, that it is not
necessary to model the details of hadronization
to parameterize leading corrections to perturbation theory,
and a hope, that plausible parameterizations
inspired by perturbation theory will lead to useful
insights at the perturbative-nonperturbative interface \cite{webberlect}.  The
hypothesis seems to be correct; whether the hope will
be realized remains to be seen, but there are preliminary
indications that it might be \cite{irr}.

How does perturbation theory imply nonperturbation corrections?
In the calculation of any IR safe quantity at NLO, we always 
encounter integrals of the general form
\be
I(\alpha,p)=f(x)\; \alpha_s(Q^2)\; \int_0^Q dk\; k^p,\
\label{Ias}
\ee
with $p>-1$, where $k$ may be thought of as a gluon momentum scale, and  $f(x)$ 
represents the remaining (IR finite) dependence.
Now in many cases (see Eq.\ (\ref{thrustresum}) for
example) we can argue (or derive) that higher order
corrections modify (\ref{Ias}) to
\be
I^{(\rm resum)}(\alpha,p) =f(x)\int_0^Q dk\; \as(k^2)\; k^p \dots\, .
\label{Irunas}
\ee
This result shows that the perturbative reexpansion in terms of $\alpha_s(Q)$
is asymptotic, with high orders that grow as $n!$ at large $n$.  This information
is encoded in the singularity of the perturbative expression for $\alpha_s(k^2)$
at $k=\Lambda_{\rm QCD}$.  We can further reinterpret this behavior by means of
a Borel transform, but the inverse transform will
not be unique in any case.   Taking a more practical approach, we retain the
perturbative factor $f(x)$ and the perturbative range in the $k$ integral in
Eq.\ (\ref{Irunas}), and simply replace the lower end of the integral,
$k<\mu_1$, for some fixed $\mu_1$, by a
parameter $\alpha_p(\mu_1)$, 
\be
 \int_0^{\mu_1} dk\; \as(k^2)\; k^p
\to \mu_1^{p+1}\, \alpha_p(\mu_1)\, ,
\label{irreplace}
\ee
Since the overall integral in  Eq.\ (\ref{Ias})
behaves as $Q^{p+1}$ this is automatically a power
correction.  Clearly the value of this approach depends 
on the assumption
that $\alpha_p(\mu_1)$ is in some sense universal \cite{irr}.
Evidently, this is true approximately, and this method has 
found applications in 
models for DIS higher twist \cite{disht}.  
  As the notation suggests, the
parameter $\alpha_p$ is often thought of as a reflection
of a universal, nonperturbative low-scale running coupling.  This is
suggested by (\ref{Irunas}) above, where it is a
moment of the lowest-order running $\alpha_s(k^2)$.  It has been argued
that the relation is more general, and that higher orders of $\alpha(k^2)$
may be incorporated into a reconstructed
effective coupling, defined through dispersion relations  \cite{dispersive,milano}.  

In interpreting these developments, it is important to keep in mind
that the values of higher-twist parameters cannot be defined
independently of perturbation theory \cite{irr}, and that 
they will change as new orders, or resummations, are computed.
A striking example of this effect was illustrated by the
NNLO analysis of \cite{nnlodis}, which reduced the size of higher
twist contributions, relative to those found in fits based on NLO.

\begin{figure}[ht]
\centerline{\epsfysize=8cm\epsfxsize=9cm
\epsffile{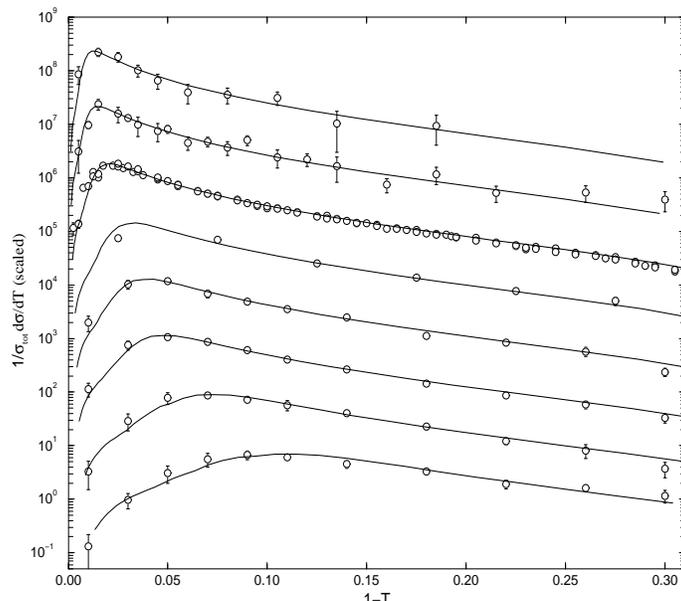}}
\caption{The comparison of thrust data with a fit at $Q=M_Z$
based on Eq.\ ({\protect\ref{shapeeq}}), at energies (from bottom to top):
$Q=14,22,35,44,55,91,133,161\; {\rm GeV}$.}.
\label{thrustfig}
\end{figure}

Another interesting application of these ideas is to resummed event
shapes, as in Eq.\ (\ref{thrustresum}), where a replacement like Eq.\ (\ref{irreplace}) leads to a
simple shift \cite{shift} in the thrust ($T=1-t$) distribution, which vanishes
as $1/Q$,
\be
{d\sigma_\pt(t) \over
dt} \to {d\sigma_\pt(t-\lambda/Q) \over dt}
+{\cal O}\left ({1\over (tQ)^2}\right)\, ,
\label{thrustshift}
\ee
with $\lambda$ a constant related to $\alpha_0$ in (\ref{irreplace}).
In a somewhat more general approach, we may once again
factorize soft gluon emission into the region between the two
jets, and derive a convolution expression for the cross section \cite{shape},
\be
{d\sigma_\pt(t) \over
dt} \to
\int_0^{tQ} d\epsilon\; f(\epsilon)\; {d\sigma_\pt(t-\epsilon/Q) \over
dt} +{\cal O}\left({1\over tQ^2}\right)\, ,
\label{shapeeq}
\ee
where
$f(\epsilon)$, a ``shape function", has 
a field-theoretic interpretation \cite{shape,ftshape}, which involves
the composite operator of Eq.\ (\ref{Wdef}).  
$f(\epsilon)$ is $Q$-independent and summarizes all $1/(tQ)^n$
corrections, while
Eq.\ (\ref{shapeeq})
reduces to (\ref{thrustshift}) with the replacement
$f(\epsilon)=\delta(\epsilon-\lambda)$. A fit \cite{shape}
to $f(\epsilon)$ using the extensive thrust data at $Q=M_Z$, and the perturbative
resummation of \cite{cttw}, faithfully predicts $d\sigma/dT$
for a wide range of $Q$, as shown in Fig.\ \ref{thrustfig}.
Given the discussion of the foregoing
section, by following this line of reasoning we may hope to relate
event shapes in $\e^+\e^-$ annihilation to energy flow
in hadronic hard-scattering cross sections.  This relation
remains unexplored, although it is certainly related to
multiplicity and correlation studies of the
final states in jet events \cite{giacomelli}.

\section{QCD at High Temperature and Baryon Number}

I have already mentioned the path from perturbative
QCD to high parton density through BFKL evolution.
Certainly, these considerations are made more interesting
by the pending turn-on of the RHIC accelerator at Brookhaven,
at which nuclei will be collided at unprecented energies. 
This development has led to a fresh look at QCD in
``extreme" conditions, long of relevance to studies of the early
universe.

{\it Color Superconductivity.}
At high enough density and temperature, the long-distance 
interactions
that lead to confinement in the normal,
hadronic phase of QCD are screened, and in some sense
nonsinglet degrees of freedom are freed.    
In fact, QCD is expected to have a possibly quite
rich phase structure in the plane of temperature (T) and
baryon density (B).

An exciting exploration of these features of the theory is the fresh look at the
 long-standing conjecture  of color superconductivity at large
B and low T, in the light of the instanton liquid model 
referred to in Sec.\ IIID.  It was
realized in Ref.\ \cite{supercond} that the four-fermion effective
Lagrange density in Eq.\ (\ref{fourquark}) produces an attractive
potential between quarks,
which can lead to a condensate of quark Cooper pairs at the Fermi surface,
in a manner analogous to 
superconductors of electric current. In fact, the condensate
can be driven by gluon exchange, but the energy gaps produced
by instantons are much larger in most, but not all \cite{son},
of parameter space.  Although more likely  to be
relevant for neutron stars than nuclear
collisions, these interesting considerations were clearly inspired by
RHIC physics, which has led to an efflorescence of studies
of the QCD B/T plane \cite{tricrit,contin}.

{\it Energy Loss.}
As a final example, I will very briefly refer to
some recent considerations on a topic of 
direct interest to RHIC and hadron-nucleus scattering, energy loss of fast partons in dense media.
High-energy partons travelling through matter (partons or hadrons)
will scatter and radiate, and their evolution into the final state
will be modified in some way.  In sufficiently inclusive processes,
these effects are (perhaps surprisingly) higher twist \cite{guo}.
It is of interest, however, also to look at changes
in radiation at transverse momentum scales set by the medium,
rather than a hard scattering.
Along these lines, recent
work on the energy loss \cite{energyloss} experienced by a quark travelling 
through a medium over length $(L)$ has identified a QCD analog
of the famous Landau-Pomeranchuk effect in QED.
This work analyzes ``induced" gluon radiation at $\L\ll k_T\ll \langle Q\rangle$,
where $\langle Q\rangle$ is the typical momentum transfer in a projectile-medium
scattering.  If the Debye length of the medium is short enough in a hot,
dense medium, $Q$ could be perturbative.  

The amplitude for an emission $q\rightarrow q+g$ at impact
parameter $b$ is denoted $f(b,t)$.  The effect of the scatterings
is a diffusion in $b$,
\be
f(b,t) \sim b\; \left\langle Q^2{d\sigma\over dQ^2}\right\rangle\; 
\exp\left[ -i\; {\rm const}\; 
\left\langle Q^2{d\sigma\over dQ^2}\right\rangle b^2t+\dots\right]\, ,
\label{fbt}
\ee
with known corrections to the Gaussion.  For large $\langle Q\rangle$,
the cross section involves the interference between
amplitudes where the radiation occurs at different positions
along the path through the medium, leading to a power spectrum
in path length and frequency given by
\be
\omega{dI\over d\o\; dz} \sim \alpha_s\, 
\int_0^L dt\; \left(1-{t\over L}\right)\, \int d^2b\; f(b,t)\; f(b,0)\, .
\ee
This results in an energy loss per unit length that grows with length (!):  
\be
-{dE\over dz} \sim {\alpha_s N_cL\over 4\lambda}\; 
\left\langle Q^2{d\sigma\over dQ^2}\right\rangle\, ,
\ee
at mean free path $\lambda$,
which is a 
diagnostic for $\left\langle Q^2 {d\sigma\over dQ^2}\right\rangle$,
potentially able to distinguish the composition of the  medium,
whether hadrons, 
Debye-screened plasma or something else.  Here again, the transition
from energy loss at relatively low transverse
momentum, to more inclusive cross sections, should be an interesting one \cite{jetloss}.

\section{Conclusion}

The central conclusion of this little review is the variety and vitality of
the work itself.  Beyond this, the nature of our knowledge
of QCD is such that important ideas and techniques only
require the possibility that they can be tested to
receive further theoretical development.  The ongoing
experiments at LEP, HERA and the Tevatron have already transformed
perturbative QCD into a truely quantitative discipline.  The
pending RHIC accelerator has inspired creative theoretical
developments.  
A strong QCD component to future high-energy projects is sure to 
be richly rewarded by insights into quantum field theory .

\section*{Acknowledgements}

I would like to thank the organizers of DPF 99 for the invitation
to participate in a vibrant meeting.  I am indebted to Werner
Vogelsang for invaluable help.
This work was supported in part by the National Science Foundation,
under grant PHY9722101.

\end{document}